\documentclass[12pt,a4paper]{article}
\usepackage[dvips]{graphics}
\usepackage{amssymb}
\begin{document}

\begin{center}
  {\large\bf Generalized Optimal Velocity Model for Traffic Flow}
\end{center}

\vspace{5mm}
\begin{center}
  Shiro Sawada\footnote{e-mail: sawada@dt.takuma-ct.ac.jp}
\end{center}

\begin{center}
{\it  Department of Telecommunications, \\Takuma National
College of Technology,\\ Mitoyo, Kagawa 769-1192, Japan}
\end{center}

\vspace{30mm}

\begin{center}
{\bf  Abstract}
\end{center}
A generalized optimal velocity model is analyzed, where
the optimal velocity function depends not 
only on the headway of each car but also the headway of 
the immediately preceding one. 
The stability condition of the model is derived by
considering a small perturbation around the homogeneous
flow solution.
The effect of the generalized optimal velocity function 
is also confirmed with
numerical simulation, by examining the hysteresis loop in
the headway-velocity phase space,
and the relation between flow and density of cars.
In the model with a specific parameter choice,
it is found that an intermediate state 
appears for the movement of cars,
where the car keeps a certain velocity though
the headway is short or long.
This phenomenon is different from the ordinary stop-and-go state.

\newpage
%%%%%%%%%%%%%%%%%%%%%%%%%%%%%%%%%%%%%%%%%%%%%%%%%%%%%%%%%%%%%%%%%%%%%%%%%%
\section{Introduction}

Traffic flow problem has been extensively studied
from physical point of view.
Fluid-dynamical model\cite{fluid}, cellular automaton model
\cite{cellular},
and the car-following model\cite{car} have been proposed and 
analyzed in detail 
to understand the mechanism of the traffic congestion on a
freeway.

Toward a realistic model which explains the traffic flow dynamics, 
the optimal velocity (OV) model proposed by Bando, Hasebe,
Nakayama, Shibata, and Sugiyama\cite{ov, ov2}
has attracted considerable interest.
Based on the second-order differential equations,
the model reveals the density pattern formation of
the congested flow of traffic without introducing 
a time lag caused by the driver's response.

Although the OV model is shown to have a universal structure
in spatio-temporal patterns in the congestion, 
most of the analyses of the model have been done in the case
where the optimal velocity function depends only on 
the headway of each car. One of the approaches to generalize the OV 
model is that 
the backward reference function is introduced\cite{hn}.
Another approach to extend the OV model is to take into
account the next-nearest-neighbor interaction\cite{nagatani},
where the optimal velocity function depends not only on
the headway of each car but also on the headway of 
the immediately preceding one.
The generalized optimal velocity function is determined
by taking into account the driver's skill, experience, 
and psychological effect, so that it is expected to
describe more realistic traffic flow.

The purpose of this paper is to analyze the generalized 
optimal velocity model proposed by Nagatani\cite{nagatani,naga}.
This paper is organized as follows.
In Section 2 the generalized optimal velocity model is reviewed and
its stability condition around the homogeneous flow solution is derived
without long-wavelength approximation.
In Section 3 numerical simulations are carried out, in particular,
the hysteresis loop in the phase space and the flow-density
relation are examined.
In Section 4 we reconsider our model to compare the result
with the one in the original model.
In the model with a specific parameter choice,
it is found that an intermediate state appears for the
movement of cars,
which is different from the ordinary stop-and-go state.

\section{Generalized Optimal Velocity Model}
%%%%%%%%%%%%%%%%%%%%%%%%%%%%%%%%%%%%%%%%%%%%%%%%%%%%%%%%%%%%%%%%%%%%%%%%%%

We first consider a dynamical model of the traffic flow
given by
\begin{equation}
  \label{eq:aa}
  \ddot x_n(t) = a(V(\Delta x_n(t), \Delta x_{n+1}(t)) - \dot x_n(t)),
\end{equation}
where $x_n(t)$ is the position of the $n$-th 
car at time $t$, $\Delta x_n(t)=x_{n+1}(t)-x_n(t)$
represents the headway of the car,
and $a$ is the sensitivity.
Thus $\Delta x_{n+1}(t)$ is the headway of the immediately
preceding car.
Here $n=1,2,\cdots,N$ is each car number
with $N$ being the total number of vehicles.
The driver's sensitivity $a$ is assumed to be independent of $n$.
Throughout this paper, we will consider the periodic boundary
condition with respect to the coordinate $x_n$ with period $L$.

At first, let us look for an appropriate
form of the optimal velocity function
to be suitable for our purpose.
The driver sometimes pays attention to not only the headway 
but also the headway of the immediately preceding one.
If the headway of the preceding car is short,
the driver assumes that the forward driver decelerates, thus
the driver decrease the optimal velocity even though
the headway of his car is long enough.
On the other hand, 
if the headway of the preceding car is long,
the driver assumes that the forward driver accelerates, thus
the driver increase the optimal velocity even though
the headway of his car is short.

Let us look at Figure 1, which describes the original optimal
velocity function in $\Delta x_n$, $\Delta x_{n+1}$, and $V$ 
space, where the numerical values
of axes are not important here.
The appropriate form of the function which satisfies the above
requirement will be seen in Figure 2.

\begin{figure}[htbp]
  \begin{center}
  \begin{minipage}[t]{.47\textwidth}
  \scalebox{0.55}{\includegraphics{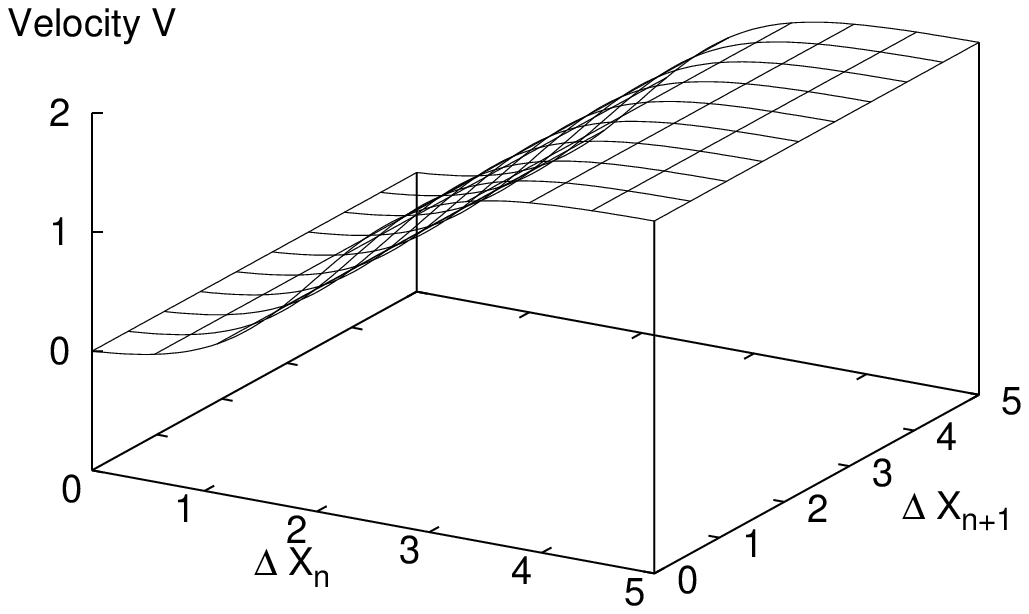}}
  \caption{The original optimal velocity function $V(\Delta x_n)$.}
  \label{fig:fig1}
  \end{minipage}
  \begin{minipage}[t]{.47\textwidth}
  \scalebox{0.55}{\includegraphics{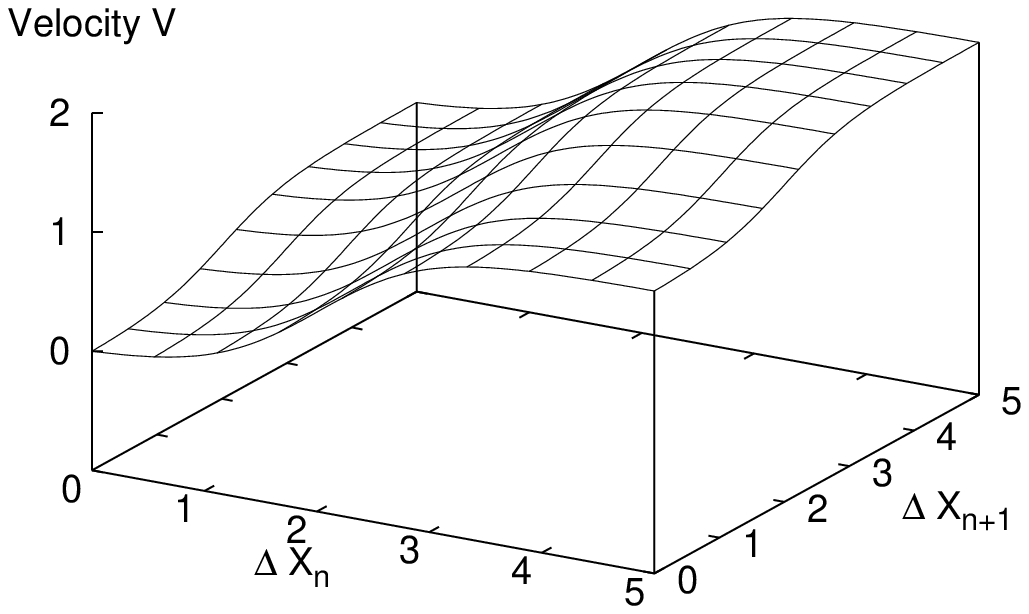}}
  \caption{The generalized optimal velocity function
    $V(\Delta x_n,\Delta x_{n+1})$.}
  \label{fig:fig2}
  \end{minipage}
  \end{center}
\end{figure}

As is already introduced in Ref.\cite{nagatani,naga}, 
we can adopt the generalized optimal
velocity function of the form
\begin{equation}
  \label{eq:bb}
  V(\Delta x_n, \Delta x_{n+1})
  =(1-p) V(\Delta x_n) + p V(\Delta x_{n+1}),
\end{equation}
where $p$ is assumed to be independent of $n$
and satisfies $0\leqq p < \frac{1}{2}$ because the dominant part of
the optimal velocity function should be $\Delta x_n$ dependent term.
The advantage of the above form is to be able to find the effect
of the additional term with varying $p$.
In spite of introducing $p$ and $\Delta x_{n+1}$ dependent term,
the above model provides exactly the same homogeneous flow 
solution as the original one without $p$ dependence.
When $p=0$, the model reduces to the original one.
According to the original OV function proposed by Bando 
et al.\cite{ov}, we will take hyperbolic tangent function
of the form
\begin{equation}
  \label{eq:cc}
  V(\Delta x_n)=\tanh(\Delta x_n-2)+\tanh(2).
\end{equation}
As we can see from eq.(\ref{eq:bb}) with eq.(\ref{eq:cc}), 
the generalized optimal
velocity function has the required properties.
In general, $p$ might depend on $n$ and also depend on time $t$.
The numerical simulation can be still performed under these circumstances.

More generally, we can consider the form
\begin{equation}
  \label{eq:dd}
  V(\Delta x_n, \Delta x_{n+1})
  =U(\Delta x_n) + W(\Delta x_{n+1}),
\end{equation}
where $U$ and $W$ have the following properties:
(i) monotonically increasing functions,
(ii) they have upper bounds, and (iii)
they satisfy $|U|> |W|$ as a realistic model.

%%%%%%%%%%%%%%%%%%%%%%%%%%%%%%%%%%%%%%%%%%%%%%%%%%%%%%%%%%%%%%%%%%%%%%%%%%
Now, let us analyze the generalized OV model given by 
eq.(\ref{eq:aa}) with eq.(\ref{eq:dd})
\begin{equation}
  \label{eq:ee}
  \ddot x_n(t) = 
  a\left(U(\Delta x_n(t))+ W(\Delta x_{n+1}(t)) - \dot x_n(t)\right).
\end{equation}
As is easily seen from eq.(\ref{eq:ee}), 
they have a homogeneous flow solution
\begin{equation}
  \label{eq:ff}
  x_n^{(0)}(t) = b n + c t,
\end{equation}
where $b=L/N$ and $c=U(b)+W(b)$.
We examine the stability against a small perturbation
$y_n(t)$ around the homogeneous flow solution (\ref{eq:ff}).
Substituting
\begin{equation}
  \label{eq:gg}
  x_n(t) = x_n^{(0)}(t)+ y_n(t)
\end{equation}
into eq.(\ref{eq:ee}), the linearized equation to $y_n(t)$ is obtained as
\begin{equation}
  \label{eq:hh}
  \ddot y_n(t) =
  a\left( g\Delta y_n(t) + h\Delta y_{n+1}(t) - \dot y_n(t)\right),
\end{equation}
where $\Delta y_n(t)= y_{n+1}(t) -y_n(t)$, and
$g$ and $h$ is the derivative of $U$ and $W$ at $b$, respectively.
The solutions to eq.(\ref{eq:hh}) is given by the Fourier series as
\begin{equation}
  \label{eq:ii}
  y_{n,k}(t) = \exp(i\alpha_k n + z t),
\end{equation}
where $\alpha_k=2\pi k/N$ with $k=1,2,\cdots, N$ and
$z$ satisfies
\begin{equation}
  \label{eq:jj}
  z^2 + a z 
  -a\left(g(e^{i\alpha_k}-1)+h(e^{2i\alpha_k}-e^{i\alpha_k})\right) =0.
\end{equation}
The stability condition is to find the condition
Re$z<0$ for all modes $\alpha_k$.
After a short calculation, it is equivalent to find the relation
\begin{equation}
  \label{eq:kk}
  F(Y)\equiv 
32h^2Y^3+16h(g-3h)Y^2+2\left((g-3h)^2-2ah\right)Y-a(g-h)<0
\end{equation}
holds for all modes $\alpha_k$, where 
we put $Y=\cos^2\frac{\alpha_k}{2}$.

In order to proceed the analysis of stability condition,
we will take the optimal velocity function adopted in
eq.(\ref{eq:bb}). 
Eq.(\ref{eq:kk}) is rewritten by
\begin{equation}
  \label{eq:ll}
  F(Y)=
  32p^2f^2Y^3+16p(1-4p)f^2Y^2
  +2( (1-4p)^2f^2-2apf)Y -a(1-2p)f <0,
\end{equation}
where $f=V'(b)$.
We can finally find that if the condition
\begin{equation}
  \label{eq:mm}
  f<\frac{a}{2}(1+2p) \quad {\rm and} \quad p\leqq\frac{1}{2}
\end{equation}
is satisfied, eq.(\ref{eq:ll}) holds for all modes $\alpha_k$.
It should be emphasized that the stability condition (\ref{eq:mm})
is derived without long-wavelength approximation. 
In the long-wavelength approximation, only the former condition
in eq.(\ref{eq:mm}) is derived from the stability analysis.
The uniform solution is unstable if $f>\frac{a}{2}(1+2p)$ or 
$p>\frac{1}{2}$.
Comparing the result with the original OV model,
we can conclude that 
the model is stabilized in
the region $ \frac{a}{2}\leqq f< \frac{a}{2}(1+2p)$
by the effect of introducing the headway of the preceding car.

%%%%%%%%%%%%%%%%%%%%%%%%%%%%%%%%%%%%%%%%%%%%%%%%%%%%%%%%%%%%%%%%%%%%%%%%%%
\section{Numerical Simulations}

To convince the analysis of stability condition for the
generalized OV model, we will
now solve
\begin{equation}
  \label{eq:nn}
  \ddot x_n(t) = a\left((1-p) V(\Delta x_n) 
    + p V(\Delta x_{n+1})- \dot x_n(t)\right)
\end{equation}
numerically.
In the simulation, $a=1$ is taken throughout this paper.
As the density of cars ($\rho=N/L$) varies and if
$f>\frac{a}{2}(1+2p)$ is satisfied,
a homogeneous flow becomes unstable and
makes a phase transition from free flow to congested one.
Looking at the spatio-temporal pattern, the congested patters 
propagate backward.
These characteristic features in the generalized model are 
the same as those appear in the original OV model.

One of the typical features of the model is that 
the movement of the car becomes the stop-and-go states in the 
congested region and the congested pattern is very stable.
It is well understood by examining the hysteresis loop in the
headway-velocity phase space.
In numerical simulations, we take
$N=100$ and $L=200$ as an example, where $f$ takes maximum value,
because the congested region is of our interest and the
result depends only on the density but not on the number of cars
and the circuit length.
The initial condition we considered here is the homogeneous
flow with small fluctuation, i.e.
\begin{equation}
  \label{eq:oo}
  x_n(0) = b n + y_n(0), \dot x_n(0) = c,
\end{equation}
where $y_n(0)$ is taken to be an uniform random 
distribution between $-0.5$ and $0.5$.

%%% figure 3.  hysteresis loop in phase space
\begin{figure}[htbp]
  \begin{center}
  \scalebox{0.7}{\includegraphics{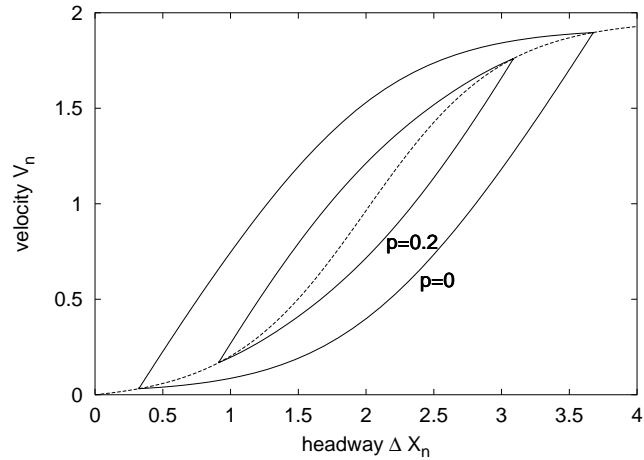}}
  \caption{The orbit of a car in the 
headway-velocity phase space after the organization of
the congestion for $p=0$ and $p=0.2$. The dashed curved line 
denotes the optimal velocity function.}
  \label{fig:fig3}
  \end{center}
\end{figure}

Figure \ref{fig:fig3} shows the orbit of a particular car in the
headway-velocity phase space with 
the parameter $p=0,~ 0.2$ as an example. 
After about 1,000 time when the generation of
the congestion is finished, the shape of the hysteresis 
loop obtained in Figure 3 never changes.
Furthermore, the shape does not depend on the initial random 
distributions.

The numerical result shows that the effect of increasing $p$ 
seems to be equivalent to increasing the sensitivity $a$.
Of course, this is also guessed from the analysis
of the stability condition in eq.(\ref{eq:mm}).
However, as will be seen in Section 4, we will find 
that the change of 
the value of $p$ can not be compensated by rescaling the 
sensitivity $a$.

Examining the bottom end point $(\Delta x_{\rm c}, v_{\rm c})
=(\Delta x_n, \dot x_n)$
and the top end point $(\Delta x_{\rm f}, v_{\rm f})=
(\Delta x_n, \dot x_n)$
in the hysteresis loop, we can obtain
the backward velocity of the congestion given by
\begin{equation}
  \label{eq:pp}
  V_{\rm back}= \frac{v_{\rm f}\Delta x_{\rm c} - 
v_{\rm c} \Delta x_{\rm f}}{\Delta x_{\rm f}-\Delta x_{\rm c}}.
\end{equation}
Numerical simulation shows that
the backward velocity of the congestion increases as
$p$ increases. 
We have also confirmed by simulating spatio-temporal patterns.
The numerical results are listed in Table 1.

\begin{table}[htbp]
  \begin{center}
\begin{tabular}[htbp]{|c|c|c|c|c|c|}
\hline
$p$&$\Delta x_{\rm c}$&$v_{\rm c}$&$\Delta x_{\rm f}$
&$v_{\rm f}$&$V_{\rm back}$\\
\hline
0.0 & 0.32274 & 0.03152 & 3.67726 & 1.89653 & 0.14791 \\
\hline
0.1 & 0.62051 & 0.08319 & 3.37945 & 1.84485 & 0.31302 \\
\hline
0.2 & 0.91196 & 0.16787 & 3.08804 & 1.76019 & 0.49945 \\
\hline
0.3 & 1.18567 & 0.29206 & 2.81434 & 1.63600 & 0.68632 \\
\hline
0.4 & 1.46814 & 0.47750 & 2.53275 & 1.45136 & 0.86548 \\
\hline
\end{tabular}
    \caption{Numerical data for $p=0,~0.1~,0.2~,0.3,~0.4.$}
    \label{tab:table1}
  \end{center}
\end{table}

%%%%%%%%%%%%%%%%%%%%%%%%%%%%%%%%%%%%%

\begin{figure}[htbp]
  \begin{center}
  \scalebox{0.9}{\includegraphics{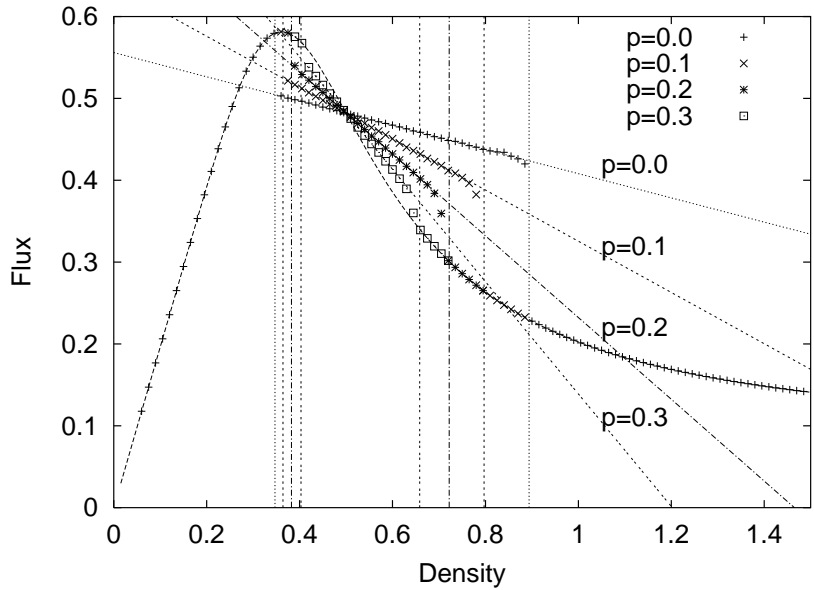}}
  \caption{Flux-density fundamental diagram
    for $p=0,~0.1~,0.2~,0.3$.}
  \label{fig:fig4}
  \end{center}
\end{figure}

Another important problem is to investigate the relation
between the flux and the density, which is called fundamental diagram.
The density $\rho$ of the cars is defined by $N/L$, where we choose
$L$=200 and vary $N$ from 10 and 300 in the simulation.
The flux $Q$ is defined by the number of cars passing by a position
per unit time.
The data was accumulated and averaged over during 20,000 time 
after first 1000 time. 
Numerical results are plotted in Figure \ref{fig:fig4}
for $p=0,~0.1,~0.2,~0.3$.
Here we omitted to plot the data which overwrites the other data.

%%% figure 4.  fundamental diagram, flow and density

In the homogeneous flow, the relation between flux $Q$
and density $\rho$ is given by
\begin{equation}
  \label{eq:qq}
  Q=\rho V\left(\frac{1}{\rho}\right)
  = \rho\left(\tanh\left(\frac{1}{\rho}-2\right) + \tanh(2)\right).
\end{equation}
In Figure \ref{fig:fig4}, it is represented by the dashed curved line.
The numerical results agree with this line in the homogeneous flow
region.

As is discussed in Ref.\cite{ov2}, the flux-density relation in the
congested flow is evaluated from the data in Table \ref{tab:table1}.
In the congested flow region, the relation between the density $\rho$
and the flux $Q$ is given by
\begin{equation}
  \label{eq:rr}
  Q = \frac{v_{\rm f}-v_{\rm c}}
{\Delta x_{\rm f}-\Delta x_{\rm c}} -V_{\rm back} \rho.
\end{equation}
Substituting the values in Table \ref{tab:table1} 
into eq.(\ref{eq:rr}), we obtain the $Q$-$\rho$ relation as in
the Table \ref{tab:table2}.
\begin{table}[htbp]
  \begin{center}
\begin{tabular}[htbp]{|c|c|}
\hline
 $p$ &  $Q$-$\rho$ relation\\
\hline
0.0 & $Q=0.55597-0.14792 ~\rho$ \\
\hline
0.1 & $Q=0.63853-0.31302 ~\rho$ \\
\hline
0.2 & $Q=0.73174-0.49945 ~\rho$ \\
\hline
0.3 & $Q=0.82518-0.68632 ~\rho$ \\
\hline
0.4 & $Q=0.91475-0.86548 ~\rho$ \\
\hline
\end{tabular}
    \caption{Flux-density relation in congested flow
      for $p=0,~0.1,~0.2,~0.3,~0.4$.}
    \label{tab:table2}
  \end{center}
\end{table}
These lines are plotted
in Figure \ref{fig:fig4} with the fundamental diagram. 
In Figure 4 the vertical lines represent the boundaries for
the stability condition given by eq.(\ref{eq:mm})
in the case of $p=0,~0.1,~0.2,~0.3$, where the same dashed
line is used as the one used to draw the predicted line
in eq.(\ref{eq:rr}).
The numerical results for various $p$ values are good
agreement with the predicted lines in Table \ref{tab:table2}.

We will summarize
the effect of the $\Delta x_{n+1}$ dependent term 
in fundamental diagram.
In the homogeneous flow region, there is no effect 
in the flux-density relation, 
because $p$ dependence in the generalized OV function disappears
in the case of homogeneous flow, as is seen in eq.(\ref{eq:bb}).
In the congested region, the flow 
increases as $p$ increases if $\rho<\frac{1}{2}$
and inversely the flow decrease as $p$ increase if $\rho>\frac{1}{2}$.
The reason is as follows.
When the density is low, the average of the 
headway is long.
If $\Delta x_{n+1}$ is long,
the larger value of the velocity
than that of the case without $\Delta x_{n+1}$ dependence
is allowed.
Hence the flow increases by taking into account the 
headway of the immediately preceding car.
Inversely, when the density is high, i.e. $\rho>\frac{1}{2}$,
the small value of the velocity is taken compared 
with the case of $p=0$,
because $\Delta x_{n+1}$ is short.
Hence the flow decreases by taking into account the 
headway of the immediately preceding car.

%%%%%%%%%%%%%%%%%%%%%%%%%%%%%%%%%%%%%%%%%%%%%%%%%%%%%%%%%%%%%%%%%%%%%%%%%%
\section{Rescaled Model}

\begin{figure}[htbp]
  \begin{center}
  \scalebox{0.7}{\includegraphics{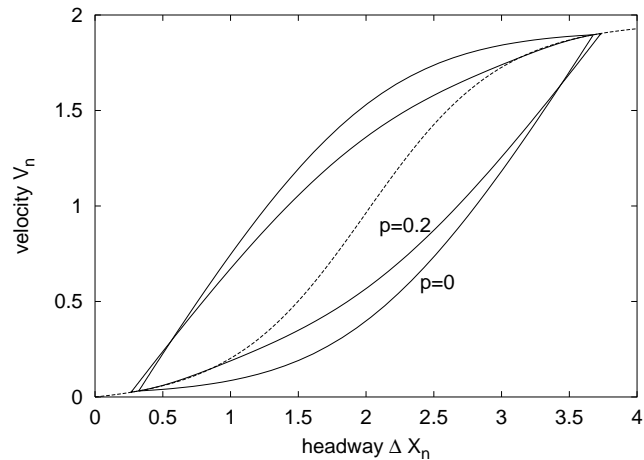}}
  \caption{Hysteresis loops for $p=0$ and $p=0.2$.}
  \label{fig:fig5}
  \end{center}
\end{figure}

As far as the stability condition of traffic flow is concerned, 
our analytical result of eq.(\ref{eq:mm}) means that the effect of
the $\Delta x_{n+1}$ dependent term can be compensated
by rescaling the sensitivity $a$.
Thus we can rewrite the generalized OV model as
\begin{equation}
  \label{eq:tt}
  \ddot x_n(t) = a\left(\frac{1-p}{1+2p} V(\Delta x_n) 
    + \frac{p}{1+2p} V(\Delta x_{n+1})- \frac{1}{1+2p}\dot x_n(t)\right).
\end{equation}
Now the stability condition of the above model is 
given by $f<\frac{a}{2}$ which is independent of $p$.
By investigating the hysteresis loop in this model with
various values of $p$, we can clarify the effect of the
headway of the preceding car.
Numerical simulation can be performed in the same way as
in Section 3.
Obtained numerical data for the headway and velocity of a particular car 
is plotted in Figure \ref{fig:fig5}.

%%% figure Hysteresis loop

We can find that the effect of the $\Delta x_{n+1}$ dependent term 
can not be compensated by rescaling the sensitivity $a$.
The effect of $p$ is now easily understood.
Compared with $p=0$, shorter value of the 
minimum of the headway is allowed,
and also longer value of the maximum of the headway 
is allowed.
Furthermore, when the car accelerates, 
larger value of the velocity can be taken even if
the headway is short.
Inversely, when the car decelerates,
smaller optimal velocity is taken even if
the headway is long enough.

The above characteristic feature in the phase space
holds for various $p$ values as long as $p<\frac{1}{2}$.
If we take $p>\frac{1}{2}$, the numerical simulation shows that
the cars take over the cars ahead, thus it is not realistic.
Of course, a room to change the form of OV function is left,
but we will not consider it here.

\begin{figure}[htbp]
\begin{center}
  \begin{minipage}[t]{.47\textwidth}
  \scalebox{0.45}{\includegraphics{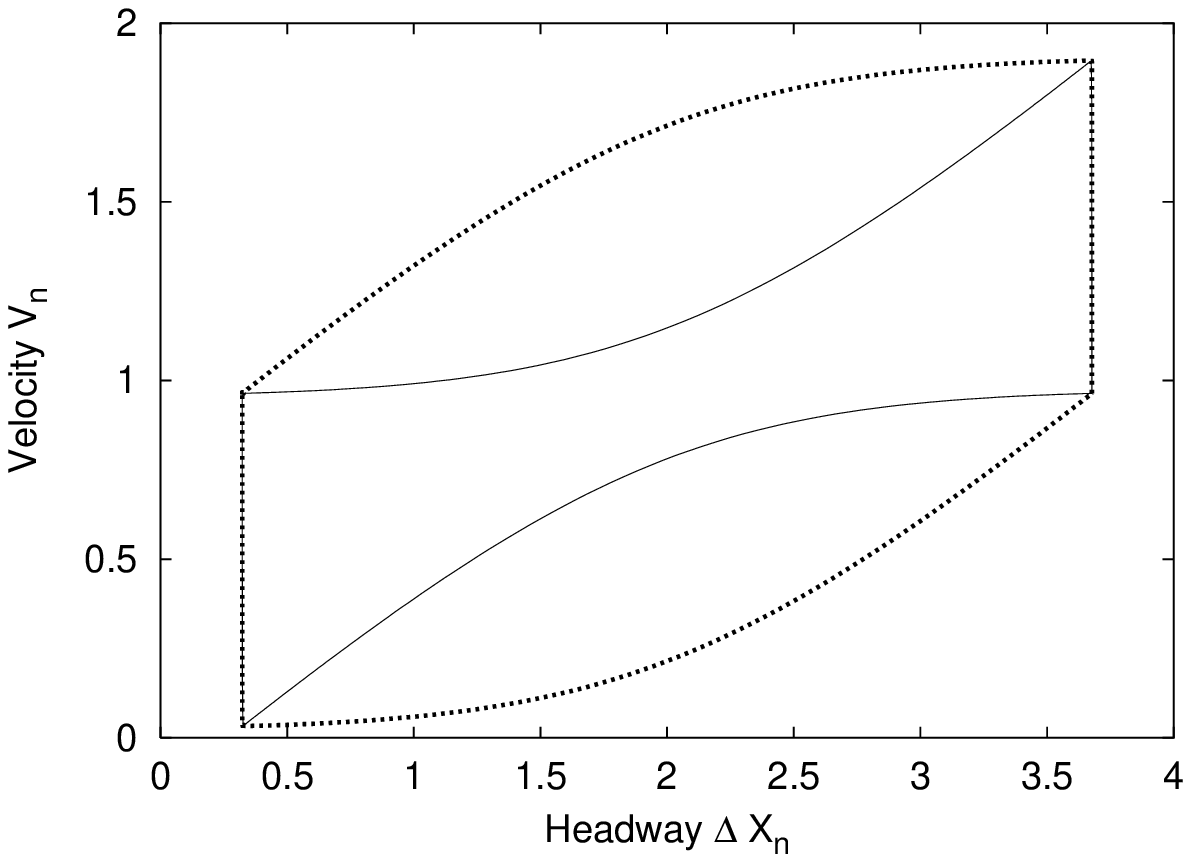}}
  \caption{Hysteresis loops for $p=\frac{1}{2}$ 
in $(\Delta x_n, \dot x_n)$ space.}
  \label{fig:p-hv}
  \end{minipage}
  \begin{minipage}[t]{.47\textwidth}
  \scalebox{0.55}{\includegraphics{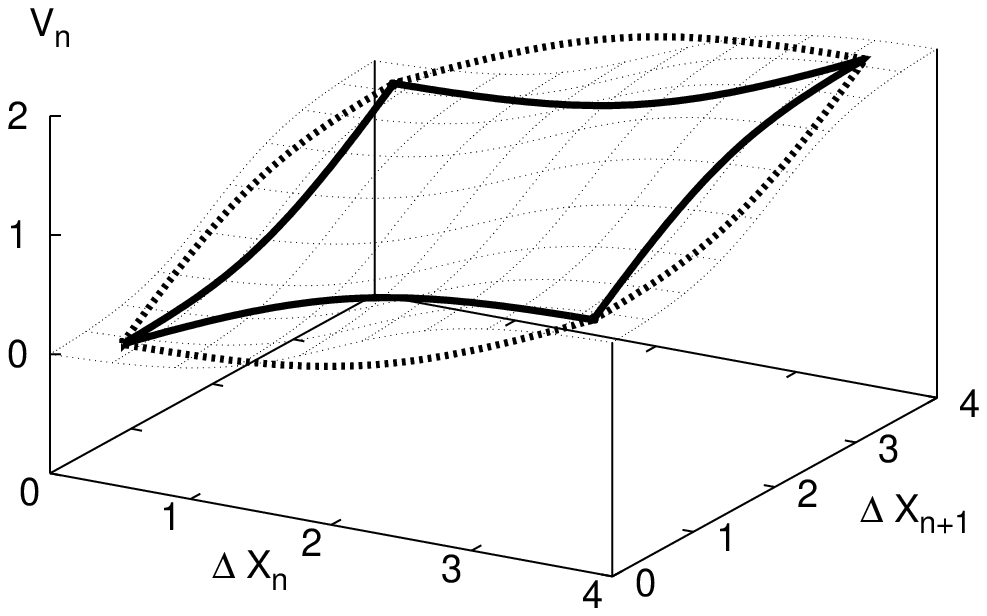}}
  \caption{Hysteresis loops for $p=\frac{1}{2}$
in $(\Delta x_n, \Delta x_{n+1}, \dot x_n)$ space.}
  \label{fig:p-hhv}
  \end{minipage}
\end{center}
\end{figure}

It is found that the model give by eq.(\ref{eq:tt})
has a different feature in
the unstable state when we take $p=\frac{1}{2}$.
Figure \ref{fig:p-hv} shows the hysteresis loop for $p=\frac{1}{2}$
after some relaxation time, e.g. about $10^8$ time in the
simulation.
More precisely, two independent hysteresis loops appear
for two successive cars.
In Figure \ref{fig:p-hv}, the line and the dotted line are 
the hysteresis loops
for the odd car number and even car number, respectively.
This hysteresis loop is well understood when the orbit of the
car is shown in the $(\Delta x_n, \Delta x_{n+1}, \dot x_n)$ phase space.
Figure \ref{fig:p-hhv} shows the orbit of two successive cars in
the phase space.
The simulation
shows that all of the vehicles sweep the same orbit as the one
in Figures \ref{fig:p-hv} and \ref{fig:p-hhv}.
We have examined the model with various initial configurations
such as random distributions, different number of cars $N$ 
and circuit length $L$.
The above result does not depend on the initial distributions
and the parameters.

\begin{figure}[htbp]
\begin{center}
  \begin{minipage}[t]{.47\textwidth}
  \scalebox{0.5}{\includegraphics{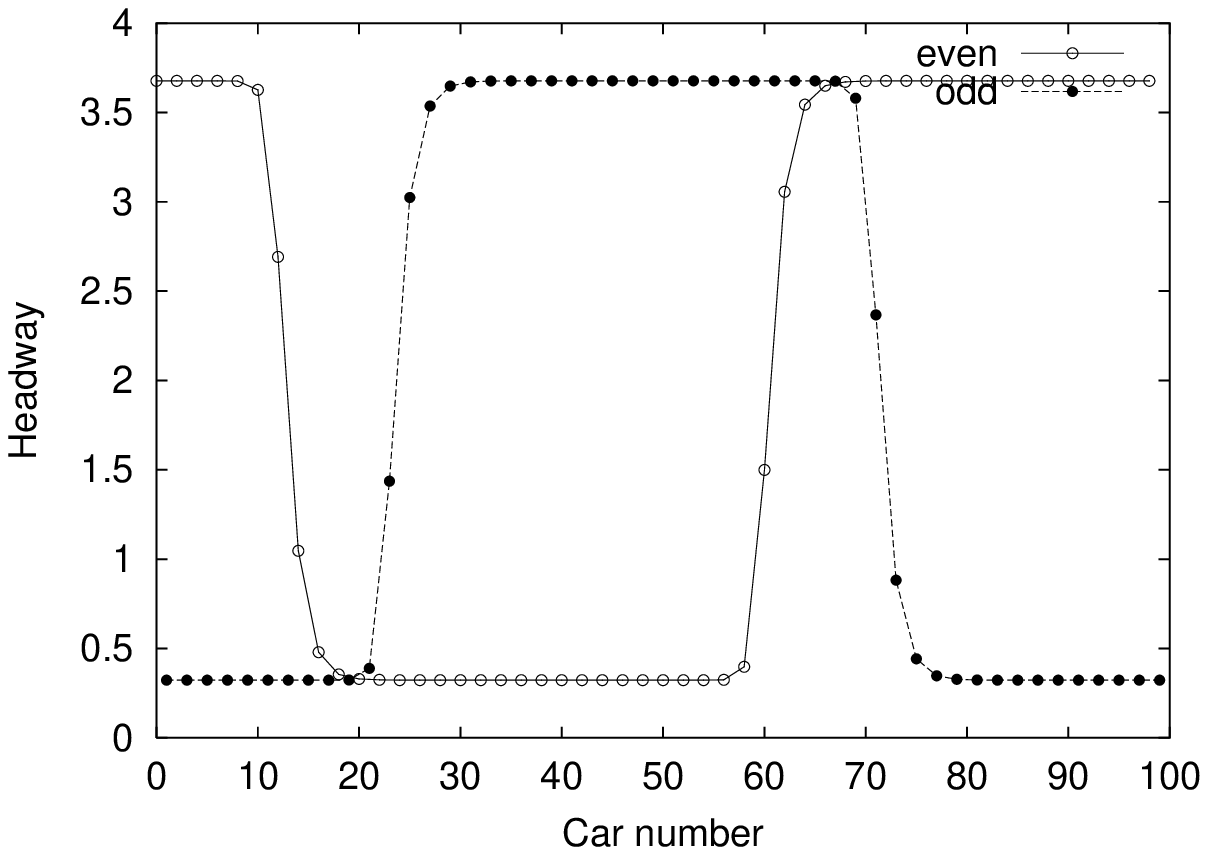}}
  \caption{Headways for all vehicles at $10^8$ time with
$N=100$, $L=200$.}
  \label{fig:p-nh}
  \end{minipage}
  \begin{minipage}[t]{.47\textwidth}
  \scalebox{0.5}{\includegraphics{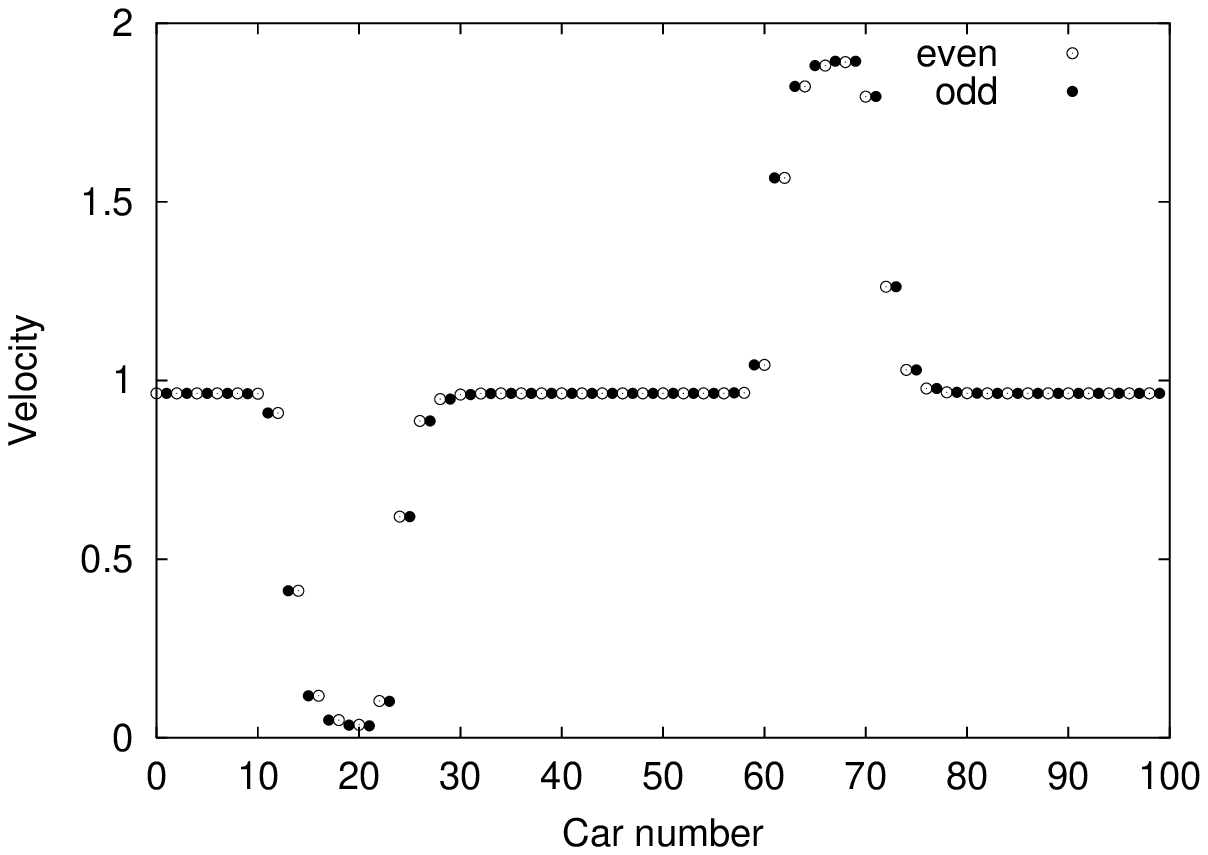}}
  \caption{Velocities for all vehicles at $10^8$ time with
$N=100$, $L=200$.}
  \label{fig:p-nv}
  \end{minipage}
\end{center}
\end{figure}

Figure \ref{fig:p-nh} and \ref{fig:p-nv} show the 
headways and the velocities for all vehicles,
respectively, in the case of $N=100$ and $L=200$ at 
$10^8$ time, starting
with arbitrary random distributions with zero velocities.
Numerical simulations show that the characteristic feature
does not depend on the initial distributions and the parameters.
%%From the numerical simulations, we can extract the common 
%%feature which appears in the model give by eq.(\ref{eq:tt})
%%with the specific parameter $p=\frac{1}{2}$.
We can find that an intermediate state appears
for the movement of cars,
which is different from the ordinary stop-and-go state.
Numerically, the state corresponds to the points
$(\Delta x_n, \Delta x_{n+1}, \dot x_n)=(3.677, ~0.323, ~0.964)$
and
$(\Delta x_n, \Delta x_{n+1}, \dot x_n)=(0.323, ~3.677, ~0.964)$
in the phase space as is given in Figure \ref{fig:p-hhv}.
Thus the intermediate state means that the car can keep a certain 
velocity (0.964) though the headway is short (0.323) or long (3.677)
with holding $\Delta x_n + \Delta x_{n+1} =4$.
The appearance of the intermediate state is universal in the
sense that it does not depend on
the initial distributions,
the number of cars $N$ and the length $L$.
It should be noted that this state never appears if $p<\frac{1}{2}$.

This characteristic feature is understood by considering the
equation for the headway which is given by
\begin{equation}
  \label{eq:vv}
\Delta \ddot x_n = \frac{a}{1+2p}\left(
(1-2p)V(\Delta x_{n+1}) + p V(\Delta x_{n+2}) -
(1-p)V(\Delta x_n) - \Delta \dot x_n\right).
\end{equation}
If $p=\frac{1}{2}$ is chosen, the first term in the
right hand side of eq.(\ref{eq:vv}) vanishes and hence
$\Delta x_{n+1}$ dependence disappears.
One might expect that the intermediate state is
understood by the nonlinear analysis near the critical
point $a=a_c$.
Following the analysis by Komatsu and Sasa\cite{ks},
the equation for the headway is derived near
the critical point and
the modified Korteweg-de Vries(MKdV) equation and its higher-oder
corrections is obtained by introducing
a small scaling parameter $\epsilon=\sqrt{(a_c-a)/a_c}$.
However, the equation near the critical point 
in the case of $p=\frac{1}{2}$ is 
exactly the same as the one in the case of $p=0$ up to
the order $\epsilon$.
Because the difference in eq.(\ref{eq:vv}) with
$p=0$ and $p=\frac{1}{2}$ near the critical point
is that the Fourier mode with $p=\frac{1}{2}$ 
is just twice that of the other one with $p=0$.
Since our analysis is performed far from the critical point,
the higher-oder corrections or non-perturbative effect
should be considered to understand the intermediate state.

%%%%%%%%%%%%%%%%%%%%%%%%%%%%%%%%%%%%%%%%%%%%%%%%%%%%%%%%%%%%%%%%%%%%%%%%%%
\section{Summary and Discussion}
We have analyzed the stability of the generalized optimal 
velocity model where the optimal velocity function
depends not only on the headway of each car but also on the
headway of the immediately preceding one.
The effect of the newly introduced $\Delta x_{n+1}$ dependent term
was examined by numerical simulation.
In particular, the hysteresis loop in the phase space
and the flux-density relation were examined in detail
by taking the various values of the parameter.

We found that the effect of the $\Delta x_{n+1}$ dependent term 
can not be compensated by rescaling the sensitivity $a$.
In the model with the specific parameter choice $p=\frac{1}{2}$,
we found that the intermediate state appears for the
movement of cars,
which is different from the ordinary stop-and-go state.
Numerical simulation shows that the appearance of
the intermediate state is universal
because it does not depend on the initial conditions.

We would expect that the model is related with the exact solution
given by Jacobi's elliptic function\cite{exact}.
It is interesting to examine the
difference-differential equation
\begin{equation}
  \label{eq:ww}
\dot x_n(t+\tau) = V(\Delta x_n(t), \Delta x_{n+1}(t)).
\end{equation}
The details of the numerical simulation and the
analysis of the underlining mathematical structure of
the generalized optimal velocity model are under study.

\begin{center}
  {\bf Acknowledgment}
\end{center}

The author would like to thank S. Tadaki for informing of
the Refs.\cite{nagatani,naga}.
The part of the numerical computation in this work was carried out at
Yukawa Institute Computer Facility.

%%%%%%%%%%%%%%%%%%%%%%%%%%%%%%%%%%%%%%%%%%%%%%%%%%%%%%%%%%%%%%%%%%%%%%%%%%
\end{document}